\documentclass[11pt,letterpaper]{article}
\usepackage[utf8]{inputenc}
\usepackage[T1]{fontenc}
\usepackage[left=1.00in, right=1.00in, top=1.00in, bottom=1.00in]{geometry}
\usepackage{graphicx}   
\usepackage{longtable}
\usepackage{lineno,hyperref}
\usepackage{amsmath, amssymb}
\usepackage{array,multirow,bigdelim}
\usepackage{xcolor}
\usepackage{ulem}
{
\begin{document}
\begin{center}
{\Large\bf Quantum Hamilton-Jacobi Quantization and Shape Invariance}
\end{center}
\vspace*{.25in} \large
\begin{center}{\large
Rathi Dasgupta\footnote{e-mail: rathi.dasgupta@gmail.com}, Asim Gangopadhyaya\footnote{e-mail: agangop@luc.edu}
}\\
Department of Physics, Loyola University Chicago, Chicago IL, USA
\end{center}

\vspace{0.15in}

\section*{abstract}
Quantum Hamilton-Jacobi quantization scheme uses the singularity structure of the potential of a quantum mechanical system to generate its eigenspectrum and eigenfunctions, and its efficacy has been demonstrated for several well known conventional potentials. Using a recent work in supersymmetric quantum mechanics, we prove that the additive shape invariance of all conventional potentials and unbroken supersymmetry are sufficient conditions for their solvability within the quantum Hamilton-Jacobi formalism.

\section*{Introduction}\label{sec.introduction}
The study of exactly solvable systems in quantum mechanics has a very rich history. One of the earliest methodical analysis of such systems was carried out by Infeld and Hull \cite{Infeld}, and their meticulous work on factorization of the Hamiltonian and the necessary condition for solvability is the first known paper on shape invariance \cite{Infeld,Miller,Gendenshtein1}. 

Supersymmetric quantum mechanics (SUSYQM) \cite{Witten,Solomonson,CooperFreedman}  is now a well established method and is widely used to study exactly solvable models.  It is a generalization of the ladder operator formalism of Dirac and F\"ock for the harmonic oscillator system where a quadratic hamiltonian is factorized into a product of two self-adjoint first order differential operators, the ladder operators ${\rm a}^\pm \equiv \left( \mp \hbar \frac{d}{dx} + \frac12 \omega x\right) $ 
\footnote{In this paper, we have set $2m=1$.}. 
Generalizing these ladder operators by $A^{\pm} \equiv \mp\, \hbar \, d/dx + W(x,a)$, where $W(x,a)$ is a real function of $x$ and a parameter $a$, we can generate a Hamiltonian $H_-=A^+A^-$. However, this Hamiltonian is in general not exactly solvable. But, if the function $W$, known as the superpotential, obeys two additional properties of unbroken supersymmetry and shape invariance \cite{Infeld,Miller,Gendenshtein1}, then the spectra for such quantum mechanical systems can be determined without solving the Schr\"odinger differential equation. 

Thus, within the SUSYQM formalism, a system is exactly solvable if it has unbroken supersymmetry (SUSY) and shape invariance (SI). Since there are many other paths to quantization \cite{Styer2002}, a question then naturally arises whether unbroken SUSY and SI render a problem exactly solvable in other quantization methods as well. We believe the answer is yes. However, given that supersymmetry and shape invariance play different roles in different formalisms, we feel it is important to show how solvability follows from them in each case. In a recent work \cite{Rasinariu2013} it was proved that shape invariance and unbroken SUSY suffices to determine the spectrum of a quantum system in deformation quantization formalism. In this paper we show that the SI and unbroken SUSY are indeed sufficient to ensure the solvability of a system within Quantum Hamilton-Jacobi quantization (QHJ) introduced by  Leacock and Padgett \cite{Leacock1,Leacock2} and  Gozzi \cite{Gozzi}. 

Quantum Hamilton-Jacobi quantization is an elegant method to determine eigenspectra for quantum mechanical systems from the singularity structure of the underlying potential. On case-by-case basis, many researchers have demonstrated the efficacy of this formalism to determine eigenvalues and eigenfunctions for a multitude of potentials \cite{Kapoor1,Kapoor2,Ganguly2006,Yesiltas2008,Ocak2008,Yesiltas2010,Gharbi2013,Girard2015,Gu2016,Poveda-Cuevas2016,Schatz2018,KapoorBook}. 
In Ref. \cite{Gangopadhyaya2007}, the authors showed that QHJ and the shape invariance condition do help determine the spectrum for superpotentials that are either algebraic or exponential functions of the coordinate in unbroken supersymmetric phase; the general form of all conventional potentials were not known at that point. In Refs. \cite{Gangopadhyaya2008,Bougie2010,symmetry}, authors showed that for all conventional superpotentials must have the form $W(x,a) = a\,f_1(x)+f_2(x) + u(a)$. In this paper, starting from the above form for superpotential and unbroken supersymmetry, we determine spectra for all conventional potentials using the QHJ formalism. Most importantly, our derivations do not use explicit forms of any potentials, instead we work with the above general form $W(x,a)$ with functions $f_1(x)$ and $f_2(x)$ satisfying additional constraints stemming from shape invariance. Furthermore, guided by Ref. \cite{Kapoor3,Girard2015}, we also show that QHJ can be used to derive the eigenfunctions directly from this general form.  

Our layout for this paper is as follows: we first introduce the QHJ formalism and briefly describe SUSYQM. In particular, we discuss additive shape invariance and list all shape invariant classes that comprise the complete set of conventional potentials.  We then determine their spectra using QHJ, without invoking the specific functional form of any of the potentials. We also derive the eigenfunctions for a class of superpotential that is a fair representative of all conventional potentials.  In our derivations of both eigenvalues and eigenfunctions, we emphasize the indispensable role the unbroken SUSY plays in Quantum Hamiton-Jacobi formalism.

 In Table \ref{table:conventionalSIP}, we list all classes of superpotentials derived from shape invariance \cite{Gangopadhyaya2008,Bougie2010,symmetry}, their energies and connect them with corresponding conventional potentials.
\section*{Introduction to QHJ and SUSYQM}\label{sec.qhj&susyqm}

\subsection*{Quantum Hamilton-Jacobi Quantization}
For one-dimensional quantum mechanical systems, QHJ \cite{Leacock1,Leacock2,Gozzi} formalism works with the quantum momentum function  (QMF) of a particle defined \footnote{Note that this definition differs from that of ref. \cite{Leacock1,Kapoor1} where it is define as $p_{\rm }(x)=-i\frac{\psi^\prime (x)}{\psi(x)}$.  } by $p(x) = - \hbar \, \psi^\prime (x)/\psi(x)$, where $\psi$ is the wave function of particle and prime denotes differentiation with respect to the coordinate $x$. Thus,  $p(x)$ has singularities at points where $\psi(x)=0$. 
Substituting $p(x)$ into the Schr\"odinger equation 
\begin{equation} -\hbar^2 \psi^{\,\prime\prime} + \left( V(x)-E\right)\psi = 0~,
	\label{SchEq1}
\end{equation}
we arrive at the quantum Hamilton-Jacobi equation
\begin{equation} 
	p\,^2(x) -\,
	\hbar \, p^{\,\prime}(x) = V(x) - E~.
	\label{QHJ1} 
\end{equation}

We will now investigate the singularities of $p(x)$ to determine eigenvalues and eigenfunctions of a quantum mechanical system with shape invariance and supersymmetry. In addition to the singularities generated by the nodes of an eigenfunction, the function $p(x)$ also inherits contribution to its singularity structure from the potential $V(x)$. 

After a brief introduction of SUSYQM, we will show that the singularity structure of  $p(x)$ for a system, along with shape invariance and unbroken SUSY, is sufficient to determine its eigenvalues and eigenfunctions.  Our work will be guided by the  Refs. \cite{Leacock1,Leacock2,Kapoor1,Kapoor2,KapoorBook,Kapoor3,Gharbi2013,Girard2015} that analyzed many specific potentials. 
  
\subsection*{Supersymmetric Quantum Mechanics}
In supersymmetric quantum mechanics, similar to the algebraic method for the harmonic oscillator pioneered by Dirac and F\"ock, a Hamiltonian is written as a product of two linear differential operators $A^{\pm} \equiv \mp\, \hbar \, d/dx + W(x,a)$ that are hermitian conjugates of each other. The function $W(x,a)$ is known as the superpotential.  Products of these operators  generate two partner Hamiltonians $H_{\mp}$ given by
\begin{equation}
A^{\pm}A^{\mp}~=~H_\mp ~=~ - \hbar^2  \frac{d^2}{dx^2} +  W^2(x)\mp\hbar\, W^\prime(x)~,
\end{equation}
with  $V_\mp(x)= W^2(x)\mp \hbar \,W^\prime(x)$ as their potentials. 
Since these Hamitonians are positive definite operators, their eigenenergies must be greater than or equal to zero. Furthermore, it can also be shown that both $H_{\mp}$ cannot have a zero-energy groundstate. If the groundstate energy of one of the Hamiltonians is zero, we say that the supersymmetry is unbroken.  In such a case the groundstate energy of the partner Hamiltonian must be positive.

These Hamiltonians  $H_{\mp}$ are related by the intertwining conditions 
$$A^-H_- = H_+A^- \qquad \mbox{and} \qquad  A^+H_+ = H_-A^+~,$$
which, for the unbroken supersymmetric case, lead to the isospectral relations
\begin{equation}
	\frac{~~~{A}^- }{\sqrt{E^{+}_{n} }} ~\psi^{(-)}_{n+1} 
	= ~\psi^{(+)}_{n} ~~; ~~ ~~
	\frac{~~~{A}^+}{\sqrt{E^{+}_{n} }}~\psi^{(+)}_{n} 
	= ~  \psi^{(-)}_{n+1}~, \label{eq.isospectrality1}
\end{equation}
and
\begin{equation}
	E_{n+1}^{(-)} = E_{n}^{(+)} , \quad ~ n=0,1,2,\cdots~, \label{eq.isospectrality2}
\end{equation}
where $E_{n}^{(\pm)}$ and $\psi^{(\pm)}_{n}$ are respectively the eigenvalues amd eigenfunctions of $H_\pm$. We restrict our analysis to systems with unbroken SUSY \footnote{One of the main reasons for not considering the conventional potentials with broken SUSY is that a very small number of  them can hold boundstates \cite{Dutt1993,Gangopadhyaya2001,Gangopadhyaya2021}. Furthermore, the potentials that can hold boundstates are exactly those that can be mapped to unbroken supersymmetric phase with special discrete transformations \cite{Gangopadhyaya2021}.
}, 
and choose the Hamiltonian $H_- = {A}^+ {A}^-$ to have a zero energy ground state, and hence have $E_{0}^{(-)}=0$. Thus, we have 
${A}^+ {A}^- \psi_0^{(-)}(x) =0$, which implies
\footnote{Since $\left( {A}^-\right) ^\dagger = {A}^+$, the equation ${A}^+ {A}^- \psi_0^{(-)}(x) =0$ implies $\int |{A}^- \psi_0^{(-)}(x)|^2 dx =0$, and hence ${A}^- \psi_0^{(-)}(x) =0$.} 
${A}^- \psi_0^{(-)}(x) =0$, and leads to the zero-energy groundstate:		
\begin{equation} \label{eq.groundstate}
\psi_0^{(-)}(x,a) = {\mathcal N}\, e^{-\frac1\hbar \int_{x_0}^x W(x,a) dx}~,
\end{equation}
where $ {\mathcal N}$ is the normalization constant 
\footnote{ For this groundstate to be normalizable, we need $\int_{x_0}^{\pm \infty} W(x,a) dx =0$, which requires that $W$ be negative as $x\to -\infty$ (the left boundary of the domain) and be positive as $x\to \infty$ (the right boundary).	
}
and $x_0$ is an arbitrary but finite point on the real axis.  Thus, for a system with unbroken SUSY, we see that both the eigenvalue and the eigenfunction of the groundstate are exactly known.  As we will see in a later section, unbroken SUSY plays a very important role in determining the eigenvalues and eigenfunctions of $H_-$ when the superpotential $W$ is shape invariant.

\subsection*{QHJ}
In terms of the superpotential $W$, the quantum Hamilton-Jacobi equation becomes
\begin{equation} 
	p\,^2(x) -\hbar \,p^{\,\prime}(x) = W^2(x)- \hbar W^\prime(x) - E \label{eq.QHJ2} ~.
\end{equation}
From Eq. (\ref{eq.QHJ2}), it appears that if $E$ were to be set zero, $p$ should go into the superpotential $W$, which is indeed correct.
Since the groundstate energy is zero, from the condition $A^{-}  \psi_0^{(-)}(x) = 0$, the superpotential can be written as $W(x,a) = - \hbar \, {\psi_0^{(-)}}^\prime(x) /\psi_0^{(-)}(x)$.
Comparing this expression for $W(x,a)$ with $p(x) = - \hbar \, \psi^\prime (x)/\psi(x)$,  we see that in cases with unbroken SUSY the following limit must hold:
\begin{equation}
	\lim_{E\rightarrow\, 0} ~p(x) \rightarrow W(x,a)  \label{eq.Limit_of_p}~.
\end{equation}

As we shall soon see, since Eq. (\ref{eq.QHJ2}) is quadratic in $p$, we will be faced with indeterminacy in choosing the correct signs for residues of the quantum momentum function $p(x)$. In such cases, Eq. (\ref{eq.Limit_of_p}) will play an extremely important role in resolving this indeterminacy. 

The QHJ method is centered around analyzing the pole structure of the $p=- \hbar \left( \frac{\psi^{\,\prime}}{\psi}\right) $. The $n$-th eigenfunction of one-dimensional quantum mechanical system has $n$ nodes within the classical turning points, which implies the presence of $n$ singular points for the quantum momentum function $p(x)$ on real axis. From Eq. (\ref{eq.QHJ2}), these singular points can be shown to be simple poles, each with a residue of $-\hbar$. They are known as the ``moving poles".  
\begin{figure}
	\centering
	\includegraphics[width=0.4\linewidth]{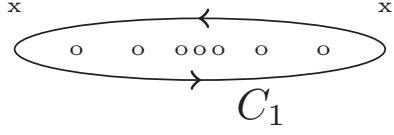}
	\caption{The diagram depicts seven moving poles that are denoted by ``o" and two fixed poles denoted by ``x". The Counter-Clockwise loop $C_1$ encircles all moving poles and excludes the two fixed poles.}
	\label{fig:ccw2cw-ccw}
\end{figure}
An integration on the complex $x$-plane along a counter-clockwise contour $C_1$ enclosing these moving poles, as shown in Fig. (\ref{fig:ccw2cw-ccw}), then lead to the following quantization condition:
\begin{eqnarray}
\oint p(x) \,dx = 2\pi i\, \left( -n \hbar\right)  ~,\label{eq.Q_Condition0}
\end{eqnarray}
where the  contour closely hugs and encloses all $n$-moving poles, and only the moving poles of the system.  

The function $p$ also has singularities that emanate from the structure of the potential $V_-(x)$  of the system which are known as the ``fixed singularities" \cite{Leacock1,Leacock2,Kapoor1,Kapoor2}. 
\begin{figure}
	\centering
	\includegraphics[width=0.5\linewidth]{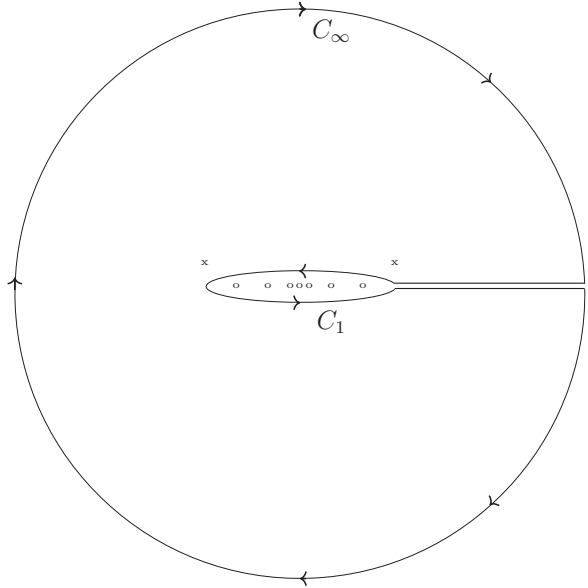}
	\caption{By cutting open one end of the loop $C_1$ of Fig. (\ref{fig:ccw2cw-ccw}) and closing it by a loop $C_\infty$ at infinity, we get a clock-wise loop that now surrounds the two fixed poles. The two parallel lines do not contribute to the integral as they are traversed in opposite directions.}
	\label{fig:ccw2cw-cw}
\end{figure}
As shown in Fig. (\ref{fig:ccw2cw-cw}), the above counter-clockwise contour enclosing the moving singularities can then also be viewed as a clockwise contour enclosing the ``fixed singularities" that were outside of the contour $C_1$ in Fig. (\ref{fig:ccw2cw-ccw}).  Now, reversing the orientation of the integration on loop $C_\infty$ to CCW, we can express the quantization condition as
\begin{eqnarray}
\left. \oint p(x) \,dx \right|_{\rm fixed~poles}= + 2\pi i\, n \hbar ~~, \label{eq.Q_Condition}
\end{eqnarray}
where the difference in sign between Eqs. (\ref{eq.Q_Condition0}) and  (\ref{eq.Q_Condition}) is due to the change in direction in which the integration is carried out. 

The complex $x$-plane with a point at infinity added can be folded into a Riemann sphere. 
The above change of orientation of the loop can also be seen as a move of the loop over the Riemannian sphere as shown in Fig. (\ref{fig:ccw2cw-cw-Riemann}). 

The Eq. (\ref{eq.Q_Condition}) is the core of much of this paper as it will help us determine the eigenvalues of all conventional potentials. 
\begin{figure}
	\centering
	\includegraphics[width=0.4\linewidth]{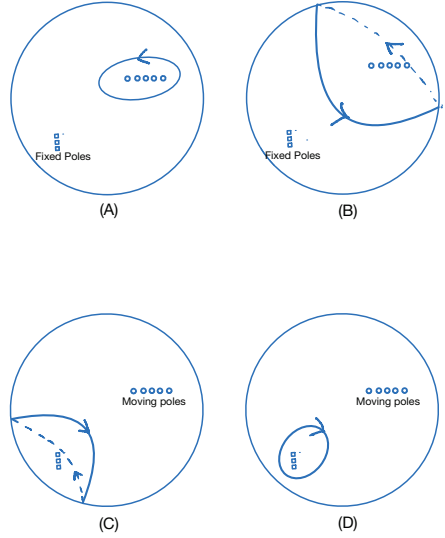}
	\caption{The subfigure (A) shows a CCW loop around the moving poles on the Riemannian sphere. The top of the loop is then pulled behind the sphere as shown in (B) and (C), and finally it emerges as a CW loop around fixed poles as shown in (D).}
	\label{fig:riemannsphere}
\end{figure}
\subsection*{Shape Invariance and Conventional Potentials}\label{sec.shapeinvariance}
In SUSYQM, a system is exactly solvable; i.e., its eigenvalues and eigenfunctions can be explicitly written in terms of the parameters of the system \cite{Cooper-Khare-Sukhatme,Gangopadhyaya-Mallow-Rasinariu} if the system has unbroken supersymmetry and the superpotential satisfies the following Shape Invariance condition \cite{Infeld,Miller,Gendenshtein1}:
\begin{equation} 
W^2(x, a_0) + \hbar \frac{d\,W(x, a_0)}{dx} +g(a_0) = W^2(x, a_1 ) - \hbar 
\frac{d\,W(x, a_1 )}{dx}+g( a_1 ).\label{eq.SIC_W}
\end{equation} 
The constants $a_0$  and  $a_1$ are parameters of the potential and are related by  $a_1 = f(a_0)$.  While we restrict our discussion in this paper to the additive shape invariance condition with  $a_1 = a_0+\hbar$, it is important to note that there are other forms of shape invariances such as multiplicative shape invariance with $a_1 = q\, a_0$  \cite{Khare1993,Barclay1993} and cyclic shape invariance  where parameters come back to the original value after a certain number of parameter changes \cite{Gangopadhyaya1996,Sukhatme1997}, which is known as its order.  For a cyclic superpotential of order $n$, we have $a_n=a_0$, and $a_n \neq a_i  ~\mbox{for all }~ i \in (1,\cdots,n-1)$.  The major reason for not including these forms of shape invariance in our analysis is that a) the multiplicative shape invariant superpotentials cannot be written in a closed form except for certain limiting values of $q$ for which they go into one of the conventional superpotentials and b) the explicit forms of the cyclic superpotentials are not known except for the case of $n=2$.  The  cyclic superpotential of order 2 leads to a harmonic oscillator potential defined over the entire $x$-axis with a \textit{soft} inverse-square singularity
\footnote{The coefficient $\alpha$ of the inverse-square singularity $\left( \frac{\alpha}{x^2}\right) $ for $n=2$ cyclic potential  necessarily falls in the transition region: $-\frac14 < \frac{\alpha}{\hbar^2}<\frac34$. This ensures that the wavefunction is normalizable at the origin and the two semi-infinite halves of the real axis can communicate with each other.}
at the origin \cite{Gangopadhyaya1994,Gangopadhyaya1996}. 

We will now show how the repeated use of Eqs. (\ref{eq.isospectrality1}), (\ref{eq.isospectrality2}) and the shape invariance Eq. (\ref{eq.SIC_W}), generate the spectrum for a system with unbroken supersymmetry. The Eq.  (\ref{eq.SIC_W}) can be written as 
$$H_+(x,a_0)+g(a_0) = H_-(x,a_1)+g(a_1)~.$$
Focusing on the groundstate of both sides, we find $E_0^{(+)}(a_0)+g(a_0) = E_0^{(-)}(a_1)+g(a_1)$. 
From the unbroken supersymmetry, we find the groundstate eigenvalues and the eigenfunction for $H_-(x,a_1)$ to be given by $E_0^{(-)}(a_1) =0$ and $\psi^{(-)}_{0}(x,a_1) = ~{\mathcal N}\, e^{-\frac1\hbar \int_{x_0}^x W(x,a_1) dx}$.
Hence from Eq. (\ref{eq.isospectrality2}) and (\ref{eq.groundstate}), we have 
\begin{equation}\label{eq.firstexcitedstate}
	E_1^{(-)}(a_0) = g(a_1)- g(a_0), ~~{\rm and}~~
	\psi^{(-)}_{1}(x,a_0) 
	=\frac{~~~{A}^-(a_0) }{\sqrt{E^{+}_{n}(a_0) }} ~{\mathcal N}\, e^{-\frac1\hbar \int_{x_0}^x W(x,a_1) dx}~.
\end{equation}
Thus, we have now derived the eigenvalues and eigenfunctions for the first excited state of the Hamiltonian $H_-(x,a_0)$. This procedure can be repeated further to determine eigenvalues and eigenfunctions for the higher excited states of $H_-(x,a_0)$, and we get
\begin{eqnarray}
		E_n^{(-)}(a_0)&=&g(a_n)-g(a_0), \label{eq:En} \\
		\nonumber\\
		\psi^{(-)}_{n}(x,a_0)&=&
		\frac{{\mathcal A}^+{(a_0)} 
			~ {\mathcal A}^+{(a_1)}  \cdots  {\mathcal A}^+{(a_{n-1})}}
		{\sqrt{E_{n}^{(-)}(a_0)\,E_{n-1}^{(-)}(a_1)\cdots E_{1}^{(-)}(a_{n-1})}}~\psi^{(-)}_0(x,a_n)
		~.
\end{eqnarray}
The derivation of $E_n^{(-)}(a_0)$ and $\psi^{(-)}_{n}(x,a_0)$ crucially depended not only on the shape invariance, but also on the unbroken supersymmetry. It allowed us to generate the spectrum by providing a bottom to the energy.  It also allowed us to determine the all excited states $\psi^{(-)}_{n}(x,a_0)$ via the groundstate wavefunction $\psi^{(-)}_0(x,a_n)=
~{\mathcal N}\, e^{-\frac1\hbar \int_{x_0}^x W(x,a_n) dx}$.

\subsubsection*{Conventional Potentials from Shape Invariance}\label{sec.shapeinvariance}
In Ref.  \cite{Gangopadhyaya2008}, the authors showed that as long as the superpotential was only a function of $x$ and $a$ and did not depend on $\hbar$ explicitly, then to the lowest power in $\hbar$, Eq. (\ref{eq.SIC_W}) would be equivalent to 
 \begin{eqnarray}
	W \, \frac{\partial W}{\partial a} - \frac{\partial W}{\partial x} + \frac12 \, \frac{d g}{d a}~=0. \label{eq.PDE1}
\end{eqnarray}
Furthermore, the authors of \cite{Bougie2010,symmetry} then showed that the inclusion of all higher order terms in $\hbar$ implies that the  superpotential $W$ would need to obey one additional partial differential equation:
\begin{eqnarray}
	\frac{\partial^{3}}{\partial a^{2}\partial x} ~W(x,a)&=& 0~,\label{eq.PDE2}
\end{eqnarray}
which then states that all conventional superpotentials must be of the form:
\begin{equation}
	W(x,a) = a f_1(x) + f_2(x)+u(a) ~. \label{eq.PDE2solution}
\end{equation}
This form was conjectured by Infeld et al. \cite{Infeld} almost seventy years ago and had been used as an ansatz by others \cite{Ramos99,Ramos00,Che2003}.

Eqs. (\ref{eq.PDE1}) and (\ref{eq.PDE2solution}) can be used to solve for the function $g(a)$, which then generates all eigenvalues for the system by $E_n=g(a+n\hbar)-g(a)$. It is worth noting that in order to avoid level-crossing, the energies must increase with $n$, which implies that we must have $\frac{d g}{d a}>0$.  This condition often determines the sign of $a$.

From Eq. (\ref{eq.PDE2solution}), we see that all conventional superpotentials fall into three classes:
\begin{enumerate}
	\item[] \textbf{Class I}: $f_1=\alpha$ is a constant;
	\item[] \textbf{Class II}: $f_2=\beta$ is a constant, and 
	\item[] \textbf{Class III}: neither $f_1$ nor $f_2$ are constants.
\end{enumerate}
These three classes comprise the complete list of all conventional superpotentials, and include all cases tabulated in Refs. \cite{Infeld} and \cite{Dutt}.  In Table. \ref{table:conventionalSIP}, we have listed these three classes, their energies and various conventional potentials they correspond to.

To determine the form of the superpotential $W$ for each of these three classes, we will need to investigate the constraints on the functions $f_1$ and $f_2 $ that are engendered by the shape invariance condition.  Substitution of $W(x,a) = a f_1(x) + f_2(x)+u(a)$  into  Eq.  (\ref{eq.PDE1})  yields
\begin{eqnarray}
	\label{eq:fourterms}
	\left(f_1f_2-f_2^\prime\right) + a \left( f_1^2-f_1^\prime\right) 
	+ f_1\left(u+a \dot{u}\right) - \dot{u}f_2 = -\left( \frac12 \,\dot{g}+\dot{u} u\right) ~,
\end{eqnarray}
where $\dot{u}$ refers to the derivative of $u$ with respect to $a$.  We will soon investigate the implication of Eq. (\ref{eq:fourterms}) on $f_1$, $f_2 $ and $u(a)$ for individual classes, and how they help determine the general form of $W$ for each case.

However, before delving into a specific class, we note that the functions $f_1$ and $f_2$ can be assumed to be linearly independent of each other.  Otherwise, if $f_2$ were of the form $f_2(x)=\nu f_1(x)$, then with a redefinition of $a\to a+\nu$,  we could fold it into the first term. Similarly, we assume that $u(a)$ does not have any linear or a constant term in $a$, as otherwise we could absorb them into $f_1(x)$ and $f_2(x)$, respectively.

In the next section, we will anlyze Eq. (\ref{eq:fourterms}) and apply QHJ quantization condition to determine the eigenspectrum for each of the three classes, and thus establish that additive shape invariance and unbroken supersymmetry are indeed sufficient conditions for QHJ to generate the eigenvalues for all conventional potentials.  Subsequently, using an example from class II, we demonstrate that  shape invariance  and unbroken supersymmetry are also sufficient conditions for QHJ to generate the eigenfunctions for the conventional potentials. 

\section*{Spectrum Generation for Conventional Potentials using QHJ}\label{sec.derivations}
In this section, we determine the spectra for the three classes discussed in the previous section. 
We will consider each class separately.
\subsection*{Class I: $f_1=\alpha$ is a constant; $W(x,a) = a \,\alpha + f_2(x)+u(a)$}
In this case Eq. (\ref{eq:fourterms}) reduces to 
\begin{eqnarray}\label{eq:fourterms.classI}
	\left(\alpha \, f_2-f_2^\prime\right)  - \dot{u}f_2 = -\left( \frac12 \,\dot{g}+\dot{u} u\right)-\left( a \alpha^2
	+ \alpha \left(u+a \dot{u}\right)\right)  ~. 
\end{eqnarray}
The left hand side (LHS) of Eq. (\ref{eq:fourterms.classI}) must be independent of $x$. As noted earlier, since ${u}$ cannot have a constant or a linear term in $a$, $\dot{u}$ must be a function of $a$, and hence $x$-dependence of the two terms on the LHS must be individually constant. Since $f_2(x) \ne {\rm constant}$, we must have $\dot{u}=0$ and $f_2^\prime =\alpha f_2 -\varepsilon$.  Thus, the general form for the superpotential of class I is $W(x,a) = a \alpha + f_2(x)$.  Before, we delve into QHJ formalism for this class, let us determine their eigenenergies using SUSYQM. From Eq. (\ref{eq.PDE1}), we have
$\frac{d g}{d a}  = 2\left( f_2^\prime  - \left( a \alpha + f_2(x) \right) \alpha \right) = -2(a\alpha^2+\varepsilon)$ 
\footnote{To avoid level-crossing, we must have $\frac{d g}{d a}>0$. This implies that for $\alpha=0$ we have $\varepsilon<0$ and for $\varepsilon=0$ we must have $a<0$.}
, which gives $g(a) = - a^2 \alpha^2 - 2\varepsilon a$ and 
\begin{equation}
	E_n (a) =  - \alpha^2\left[ (a+n\hbar)^2-a^2 \right]  - 2\varepsilon n\hbar ~.\label{eq.Energy.ClassI}
\end{equation}
We will next determine these same  energies using QHJ. The Eq. (\ref{eq.QHJ2}) for this class then reduces to 
\begin{equation}
p^2- \hbar \frac{dp}{dx}-a^2 \alpha ^2-(2 a \alpha -\alpha \hbar) \,f_2 +{E_n}-f_2^{\, 2}-\varepsilon \hbar=0~, \label{eq.QHJ.ClassI}
\end{equation}
where we have suppressed the $x$-dependence of the functions. To compute the integral $\oint p \,dx$, we will change the independent variable from $x$ to $f_2(x)$ 
\footnote{As we change variable from $x \to f_2(x)$, the function $p(x)$ should transform into a new function $\tilde{p}\left( f_2(x)\right) $.  However, we still denote QMF with $p$, as we believe that the resulting transformation of the function $p$ should be clear from the context.}, 
and the integral becomes $\oint {p} \,df_2/f_2^\prime$.  The Eq. (\ref{eq.QHJ.ClassI}) then transforms to 
\begin{eqnarray}
	{p}^2-\hbar\frac{d{p}}{df_2} \,(\alpha f_2 -\varepsilon) &-& a^2 \alpha ^2-(2 a  -\hbar) \, \alpha \,f_2 +{E_n}-f_2^{\, 2}-\varepsilon \hbar = 0~, \label{eq.QHJ.ClassI.f}
\end{eqnarray}
where  the factor $\left( \alpha f_2 -\varepsilon\right) $ appears as we use the chain rule  $dp/dx = dp/{df_2}\times {df_2}/{dx}$.
At this point, to go further we divide this subclass into two cases: $\alpha=0$ and $\alpha\neq 0$.

\subsubsection*{Class IA:  $\alpha=0$} 
For this class, we now have $~{df_2}/{dx}= -\varepsilon$, and hence Eq. (\ref{eq.QHJ.ClassI}) reduces to
\begin{equation}
	{p}^2- (-\varepsilon) \, \hbar \frac{d{p}}{df_2}+{E_n}-f_2^{\, 2}-\varepsilon \hbar  =0~. \label{eq.QHJ.ClassIA0}
\end{equation}
Since the potential function $f_2^{\, 2}+\varepsilon$ has a singularity at infinity on the complex $f_2$-plane, to determine residue at infinity we substitute  $z=1/f_2$. Replacing  $\frac{d{p}}{df_2}$ by ${d{{p}}}/{dz} \times {dz}/{df_2} = (-z^2) {d{{p}}}/{dz}$, we get
\begin{equation}
	{	{p}}^2+ \varepsilon \hbar \, (-z^2) \frac{d{{p}}}{dz}+{E_n}-\left( \frac1z\right)^{\, 2}-\varepsilon \hbar  =0~. \label{eq.QHJ.ClassI.2}
\end{equation}
To derive the residue at $z=0$, we substitute 
\[
{	{p}} = \frac{b_1}{z}+a_0+ a_1\,z + \cdots~,
\] 
which yields
\[
{a_0}^2+2 {a_0} {a_1} z+\frac{2 {a_0} {b_1}}{z}+z^2 \left({a_1}^2-{a_1} \epsilon  \hbar \right)+2 {a_1} {b_1}+\frac{{b_1}^2-1}{z^2}+{b_1} \epsilon  \hbar +{E_n}-\epsilon  \hbar =0~.
\]
Hence, we have $b_1=\pm 1,~ a_0=0$ and $a_1=-E_n/2$.  To choose the correct root for $b_1$, from Eq. (\ref{eq.Limit_of_p}) we note that in the vicinity of $z=0$ where ${E_n}<<W^2$, the function $p(x)$ must go into $W(x,a) = a \alpha + f_2(x) =  a \alpha + 1/z \approx 1/z$. Hence, we must choose $b_1=1$. With the change of variables from $x$ to $f_2$, and then from  $f_2$ to $z$, the corresponding differentials are related by $dx=df_2/(-\varepsilon) = dz/(\varepsilon\,z^2)$.  Thus, to compute the integral  $\oint {{p}}(z) dz/(\varepsilon\,z^2)= 2\pi i\, n \hbar$, we only need to include the linear term in $z$, and thus need to determine the coefficient $a_1$. Thus, substituting $p(z) = (-E_n/2)~z$, the integration yields 
\begin{equation}\label{eq.HO}
	E_n = -2 n \varepsilon \hbar = n \omega \hbar~,
\end{equation}
where we have set $ -2 \varepsilon =\omega $. As the spectrum derived in Eq. (\ref{eq.HO}) suggests, this subclass corresponds to the one-dimensional oscillator.

\subsubsection*{Class IB:  $\alpha\neq 0$}
With $ \alpha\neq 0 $, the superpotential is $W(x,a) = a \alpha + f_2(x)$.  For unbroken supersymmetry, for this case the derivative $W^\prime$, and hence $f_2^\prime$, must be positive. Since the constant $ \varepsilon$ in the equation $f_2^\prime = \alpha \, f_2- \varepsilon$ can be set equal to zero 
\footnote{The general solution of $f_2^\prime = \alpha \, f_2- \varepsilon$ can be written as $f_2^H +\varepsilon/\alpha$, where $f_2^H$ is the solution of the homogeneous equation   $f_2^\prime - \alpha \, f_2=0$. We then have $W= a \alpha + f_2^H(x)+\varepsilon/\alpha$.  
However, the term $\varepsilon/\alpha$ can be absorbed into the $a\alpha$-term  $a$ with $a\to  a+\varepsilon/\alpha^2$, and hence the superpotential $W$ reduces to $a \alpha + f_2(x)$, where $f_2$ satisfies  $f_2^\prime - \alpha \, f_2=0$, and thus effectively setting $\varepsilon=0$.},
the positivity of $W^\prime$ can be achieved, without loss of generality, by choosing $f_2$ to be negative 
\footnote{Note that $V_-=W^2-\hbar W^\prime$ is invariant under the transformation $W\to -W$ and $x\to -x$. This property can be used to choose $f_2$ to be negative without losing generality.}.
This implies $\alpha<0$, which by scaling of $x$ is then set equal to  $-1$.
The condition  $dg/da>0$ then gives $a<0$. Thus, Eq. (\ref{eq.QHJ.ClassI}) for this case becomes 
\begin{equation}
 p^2 + \hbar\, f_2\,\frac{dp}{df_2}    -a^2-f_2 (\hbar -2 a) +E_n-f_2^2 = 0~,
\end{equation}
where we have made a change of variable from $x$ to $f_2(x)$. Setting $p=\frac{{b_1}}{f_2}+{a_0}+{a_1} f_2+\cdots\,$, near  $f_2 \approx 0$ we get 
\begin{equation}
	\frac{b_1^2}{f_2^2}+\frac{(2 a_0 b_1 - b_1 \hbar)}{f_2} +a^2 + a_0^2 + 2 a_1 b_1 + E_n 
	=0~.
\end{equation}
This leads to $b_1=0$ and ${a_0}= \pm \sqrt{a^2-{E_n}}~$.  Using unbroken SUSY, we choose 
\footnote{With $\alpha=-1$ and $a<0$, $W=a \alpha + f_2\to -a$ as $f_2\to 0$.  Hence, we choose ${a_0}= \sqrt{a^2-{E_n}}\,\to -a$ as ${E_n}\to 0$. } 
$a_0 = + \sqrt{a^2-{E_n}}~$, so that $\lim_{E_n \to  0}\, p \to W$.
Since the change of variable transforms, $dx$ into  $df_2/f_2$, the non-zero contribution comes from the $a_0$-term and we do not need to determine any other coefficient.  

The potential also has a singularity at $|f_2| \rightarrow \infty$, hence we change variable to $z=1/f_2$. So, setting 
$
  {p} = \frac{b_1}{z}+a_0+ a_1\,z + \cdots~,
$
we find that near  $z \approx 0$ we must have
\[
\frac{{b_1}^2-1}{z^2} +\frac{2 a+2 {a_0} {b_1}+{b_1} \hbar
-a^2 -\hbar }{z}+{a_0}^2+ 2 {a_1} {b_1}+{E_n}~=0~,
\]
which is solved by $b_1=1$ and $a_0=-a$. With the change of variable from $x$ to $z$, the differential $dx = dz/z$, and hence only the $a_0$-term contributes again. Thus, the two singularities at $f_2=0$ and $z=0$ and Eq. (\ref{eq.Q_Condition}) give
\begin{equation}
	2\pi i \left( \sqrt{a^2-{E_n}}-a\right)  = 2\pi i \, n\hbar~, \label{eq.morse.changed}
\end{equation}
which then yields
\[
E_n = a^2 - \left( a+n\hbar\right)^2  ~,
\]
which,  with the identification $a=-A$, produces the correct energy for the Morse potential:
\[
E_n = A^2 - \left( A-n\hbar\right)^2~.
\]

\subsection*{Class II: $f_2=\beta$ is a constant; $W(x,a) = a f_1(x)+\beta+u(a)$} 
In this case Eq. (\ref{eq:fourterms}) reduces to 
\begin{eqnarray}
	\label{eq:fourterms.ClassII}
	a \left( f_1^2-f_1^\prime\right) + f_1\left(u+a \dot{u}+\beta\right) = -\left( \frac12 \,\dot{g}+\dot{u} u\right) + \beta \dot{u}~.
\end{eqnarray}
The left hand side (LHS) of Eq. (\ref{eq:fourterms.ClassII}) must be independent of $x$. Hence, either the two terms on the LHS must be individually constant, or the term proportional to $f_1$ must be linear in $a$. In the second case where $u+a \dot{u}+\beta = \gamma a$, the general solution for $u$ is $u=\frac12 \gamma a -\beta + B/a$.  Then, absorbing $\frac12 \gamma a$ in the term $a f_1(x)$, the superpotential becomes $W(x,a) = a f_1(x)+ B/a$.  The function $f_1(x)$ obeys $ f_1^\prime (x) = f_1^2(x)-\lambda$ for a constant $\lambda$
\footnote{For $\lambda \ne 0$, by scaling $f_1$ and $x$ we can set $|\lambda|=1$.}
. The first case gives exactly the same results as well, and hence the general form for the superpotential of class II is $W(x,a) = a f_1(x)+B/a$.  

Let us now determine their eigenenergies using SUSYQM. From Eq. (\ref{eq.PDE1}), we have
$\frac{d g}{d a}  = 2\left[ a f_1^\prime  - \left( a f_1(x)+B/a\right) \left( f_1(x)-B/a^2 \right)  \right] = 2(-a\lambda +2B^2/a^3)$ , which gives $g(a) = - a^2 \lambda - B^2/a^2$ and 
\begin{equation}
	E_n (a) =  - \lambda \left[ (a+n\hbar)^2 -a^2 \right]  + \left( \frac{B^2 }{a^2}- \frac{B^2 }{(a+n\hbar)^2}  \right) \hbar \label{eq.Energy.ClassII}
\end{equation}
We will now use the QHJ formalism to derive the same eigenspectrum. The Eq. (\ref{eq.QHJ2}) for this class then reduces to 
\begin{eqnarray}
p^2-\hbar  \left(f_1^2-\lambda \right)\,
 \frac{dp}{df_1} 
-\frac{B^2}{a^2}-f_1^2 \left(a^2-a \hbar \right)-a \lambda  \hbar -2 B f_1+{E_n} \label{eq.QHJ.classII}
=0.
\end{eqnarray}
There are two possibilities: $\lambda=0$ and $\lambda\neq 0$.  We first consider the case of $\lambda=0$.
\subsubsection*{Class IIA: $\lambda=0$}
For $\lambda=0$, we have $ f_1^\prime (x)= f_1^2(x)$ and Eq. (\ref{eq.QHJ.classII}) now reads
\begin{eqnarray}
	p^2-\hbar  \, f_1^2\,
	\frac{dp}{df_1} 
	-\frac{B^2}{a^2}-f_1^2 \left(a^2-a \hbar \right)-2 B f_1+{E_n} \label{eq.QHJ.classIIA}
	=0.
\end{eqnarray}
The potential is unbounded for large values of $f_1$. Furthermore, since the transformation from $x\rightarrow f_1$ sends $dx\to df_1/f_1^2$, we need to analyze the structure of $p$ near both $f_1=0$ and $f_1=\infty$. We first look for the singularity near $f_1=0$. 
Substituting 
\[
p=\frac{{b_1}}{f_1}+{a_0}+{a_1} f_1+\cdots ~
\]
in Eq. (\ref{eq.QHJ.classII}),  near  $f_1 \approx 0$ we get
\begin{eqnarray}
	\frac{{b_1}^2  }{f_1^2}
	+\frac{2 {a_0} {b_1}}{f_1}
- \frac{B^2}{a^2}	+{a_0}^2 +b_1\hbar 	+{E_n} 	=0~. \label{eq.QHJ.classIIA2}
\end{eqnarray}
Thus, we get $b_1=0$, $a_0={\sqrt{B^2-a^2 {E_n}}}/{a}$ and $a_1={a B}/{\sqrt{B^2-a^2 {E_n}}}$~. Since the differentials $dx = df_1/f_1^2$, the nonzero contribution near $f_1=0$ comes from the coefficient $a_1$.

For the singularity at infinity, we substitute $z=1/f_1$ 
\[
p^2+ \hbar \, \frac{dp}{dz} +{E_n}-\frac{B^2}{a^2}-\frac{a^2-a \hbar }{z^2}-\frac{2 B}{z} =0.
\]
Substituting 
\[p=\frac{{b_1}}{z}+{a_0}+{a_1} z+\cdots~,\] we get 
\[
\frac{-a^2+a \hbar +{b_1}^2-{b_1} \hbar }{z^2}+\frac{2 {a_0} {b_1}-2 B}{z}+{a_0}^2+2 {a_0} {a_1} z+{a_1}^2 z^2+2 {a_1} {b_1}+{a_1} \hbar -\frac{B^2}{a^2}+{E_n}=0~.
\]
This gives $b_1=a$ or $(-a+\hbar)$. Guided by Eq. (\ref{eq.Limit_of_p}), we choose  $b_1=a$
\footnote{Since near $z\approx 0$, $W=a\,f_1+B/a = \frac az +B/a \approx \frac az$, we choose $p \approx \frac az$. Hence, $b_1=a$.}  
. For the change of variable from $x\rightarrow f_1\rightarrow z$, we have $dz = -\, df_1/f_1^2 = -(f_1^2 \,dx)/f_1^2 = -\,dx$. Hence, only the $b_1$-term contributes near $z=0$.  Substituting the contributions from both singularities, we get 
\[
2\pi i \left( \frac{a B}{\sqrt{B^2-a^2 {E_n}}}- a\right) =2 \pi  i\, n \hbar~,
\]
which leads to 
\[
E_n = \frac{B^2}{a^2}-\frac{B^2}{(a+n \hbar )^2}~,
\]
which agrees with Eq. (\ref{eq.Energy.ClassII}). With $a=\ell+\hbar$ and $B=e^2/4\pi\epsilon_{\rm o}$, the eigenvalues match with the energies for the Coulomb potential. 
\subsubsection*{Class IIB: $\lambda \neq 0$}
The superpotentials for this class are of the form $W(x,a) = a f_1(x)+B/a$ with $ f_1^\prime (x)= f_1^2(x)-\lambda$. 
Since for $\lambda \neq 0$, we can set its magnitude to be equal to one.  Henceforth, we consider cases with $\lambda = \pm 1$.  

The general QHJ for class II, Eq. (\ref{eq.QHJ.classII}) now reads
\begin{eqnarray}\label{eq.QHJ.classII.lambda=1}
	p^2-\hbar  \left(f_1^2-\lambda \right)\,
	\frac{dp}{df_1} 
	-\frac{B^2}{a^2}-f_1^2 \left(a^2-a \hbar \right)-a \lambda  \hbar -2 B f_1+{E_n} 
	=0~.
\end{eqnarray}
This potential has singularity at $f_1=\infty$.  Furthermore, because the change of variable from $x$ to $f_1$ implies $dx \rightarrow df_1/f_1^\prime  = df_1/(f_1^2-\lambda) $, we also have singularities at $f_1=\pm \sqrt{\lambda }$.

We now consider the case $\lambda =1$.  Since $W^\prime = a f_1^\prime = a(f_1^2(x)-1)$ must be positive, the sign of $a$ would depend on whether $f_1^2(x)>1$ or $f_1^2(x)<1$. We first consider the case $a>0$.  
Expanding $p$ about $f_1=1$  as
\[
p=\frac{{b_1}}{f_1-1}+{a_0}+{a_1} \left(f_1-1\right)+\cdots~,
\]
and substituting in Eq. (\ref{eq.QHJ.classII.lambda=1}), we get 
\begin{eqnarray}
	&& \frac{{b_1}^2}{\left(f_1-1\right)^2}+\frac{2 {a_0} {b_1} }{f_1-1}
	-\frac{B^2}{a^2}-a^2+{a_0}^2-2 B
	+{E_n}
	=0~. \label{eq.QHJ.classIIB}
\end{eqnarray}
This gives $b_1=0$ and $a_0 = \frac{\sqrt{a^4+2 a^2 B-a^2 {E_n}+B^2}}a$. Since, $dx = \frac{df_1}{{f_1^2-1}}
=\frac{df_1}{(f_1+1)(f_1-1)} \approx \frac12 \, \frac{df_1}{f_1-1}$, only the $a_0$-term contributes.

A similar analysis around ${f_1= - 1}$, where  $dx \approx  -\, \frac12 \, df/({f_1+1})$, we find 
 $b_1=0$ and $a_0 =\frac{\sqrt{a^4-2 a^2 B-a^2 {E_n}+B^2}}{a}$, and again only the $a_0$-term contributes.
 
Now we carry out the analysis near $f_1\to \infty$ and hence set $z=1/f_1$.  We expand $p$ in the vicinity of $z=0$ as
\[
p={a_0}+{a_1} z+\frac{{b_1}}{z}\cdots ~.
\]
The Eq. (\ref{eq.QHJ.classII}) then becomes 
\[
\frac{-a^2+a \hbar +{b_1}^2-{b_1} \hbar }{z^2}+
\frac{2 {a_0} {b_1}-2 B}{z}-\frac{B^2}{a^2}-a \hbar +{a_0}^2
+2 {a_1} {b_1}+{a_1} \hbar +{b_1} \hbar +{E_n}=0~,
\]
and its solutions are $b_1=a$ and $b_1=-a+\hbar$, and we choose the first solution.

Since $dx = -dz$, it is the $b_1$-term that contributes. Thus, collecting contributions from all three singular points, from Eq. (\ref{eq.Q_Condition}) we get 
\[
2\pi\, i\, \left[ \frac{1}{2} \left(\frac{\sqrt{a^4+2 a^2 B-a^2 {E_n}+B^2}}{a}-\frac{\sqrt{a^4-2 a^2 B-a^2 {E_n}+B^2}}{a}\right)
-a\right] =2\pi \, i\,n \hbar~.
\]
Solving for $E_n$, we get 
\[
E_n = a^2- {(a+n \hbar )^2}+\frac{B^2}{a^2}-\frac{B^2}{(a+n \hbar )^2}~,
\]
which agrees with Eq. (\ref{eq.Energy.ClassII}). This analysis was carried out for $a>0$ and the energy generated above corresponds to the Eckart potential.  For negative $a$, a similar analysis yields exactly the same result; i.e., 
\begin{equation}
	E_n = a^2- {(a+n \hbar )^2}+\frac{B^2}{a^2}-\frac{B^2}{(a+n \hbar )^2}~. \label{eq.ClassIIB.alt0}
\end{equation}
Setting set $a=-A$, where $A>0$, we get 
\[
E_n = A^2- {(A-n \hbar )^2}+\frac{B^2}{A^2}-\frac{B^2}{(A-n \hbar )^2}~.
\]
This case corresponds to the hyperbolic Rosen-Morse \cite{Gangopadhyaya-Mallow-Rasinariu}.

The case for $\lambda<0$ for which $a$ is necessarily positive, we get 
\[
E_n ={(a+n \hbar )^2}-a^2+\frac{B^2}{a^2}-\frac{B^2}{(a+n \hbar )^2}~,
\]
and the corresponding potential is the trigonometric Rosen-Morse \cite{Gangopadhyaya-Mallow-Rasinariu}.
\subsection*{Class III: Neither $f_1$ nor $f_2$ is a constant}
For this class we start with the Eq. (\ref{eq:fourterms}) in its full form:
\begin{eqnarray}
	\label{eq:fourterms2}
	\left(f_1f_2-f_2^\prime\right) + a \left( f_1^2-f_1^\prime\right) 
	+ f_1\left(u+a \dot{u}\right) - \dot{u}f_2 = -\left( \frac12 \,\dot{g}+\dot{u} u\right) ~.
\end{eqnarray}
Asain, the left hand side of this equation must be independent of $x$. The $a$-dependence of the first two terms are constant and linear in $a$, respectively. Since  $f_1$ and $f_2$ are linearly independent,  the remaining two terms can contribute only if they were constant or linear in $a$. Since $\dot{u}$ cannot be a non-zero constant, to make a non-trivial contribution it must be linear in $a$. But that would make $\left(u+a \dot{u}\right)$ quadratic in $a$ and would force $f_1$ to be a constant, which would not be acceptable for this class. Hence, we must have $\dot{u}=0$ and $\left(u+a \dot{u}\right)=0$, and hence $u=0$. The general form of the superpotentials for this class is then  $W(x,a) = a f_1(x)+ f_2(x)$, with  functions $f_1(x)$ and $f_2(x)$ obeying $f_1^\prime = f_1^2 - \lambda$ and  $f_2^\prime = f_1\, f_2 - \varepsilon$, where $\lambda$ and  $\varepsilon$ are constants. 

Let us now determine the eigenenergies using SUSYQM before moving on to the QHJ formalism.  From Eq. (\ref{eq.PDE1}), we have $\frac{d g}{d a}  =2\big( af_1'+f_2'  - \left(af_1+f_2 \right)  \, f_1 \big) -2\lambda a-\varepsilon$, which gives $g(a) = -\lambda a^2 -\varepsilon a~.$
Thus, the energy $E_n$ is given by 
\begin{equation}\label{eq.Energy.ClassIII}
    E_n=g(a_0+n\hbar) - g(a_0) = -\lambda \left[\left( a+n\hbar\right)^2 - a^2\right] -\varepsilon (n\hbar) ~.
\end{equation}

As we now proceed to determine eigenvalues for potentials of class III, we will show that whenever $\lambda\ne 0$, the constant $\varepsilon$ must equal zero, and vice-versa.

\subsubsection*{Class IIIA: $\lambda=0$}
In this case,  $W(x,a) = a f_1(x)+  B/f_1(x)$
\footnote{In this case, $f_1'=f_1^2$ and $f_2'=f_1\,f_2 -\varepsilon$. The solution of the homogeneous equation $f_2'=f_1\,f_2$ is $f_2 = \alpha f_1$, and a particular solution is $f_2 = \frac12 \varepsilon/f_1$. Hence, the general form for $f_2$ is $\alpha f_1+\frac12 \varepsilon/f_1$.  The first term can be absorbed in the $a f_1$-term of $W$, and hence superpotential for this case is $W=af_1+ \frac12 \varepsilon/f_1 \equiv a f_1(x)+  B/f_1(x)$. Since $f_1'=f_1^2$ implies that $f_1\ne 0$, $f_1$ must have a definite sign. Hence, as required by unbroken SUSY,  for the superpotential $W$ to change sign we must have $\frac12 \varepsilon=B\ne 0$. }
.  
Thus,  $W(x,a)$ has singularities both at $f_1=0$ and $f_1=\infty$, which ensures that all physical moving poles will be on the positive real axis on the complex $f_1$-plane. Changing variables from $x\to f_1$, we find that the domain of the potential is semi-infinite; i.e., $f_1\in(0,\infty)$.  
The QHJ condition of Eq. (\ref{eq.QHJ2}) reduces to 
\[
p^2 -\hbar\, f_1^2\, \frac{dp}{df_1}  - \left(a^2-a \hbar \right)f_1^2-2 a B-\frac{B^2}{f_1^2}-B \hbar  +{E_n} =0.
\]
We will now expand $p(f_1)$ around the singular points at $f_1=0$ and $f_1\to \infty$.  We first consider the expansion about $f_1=0$.  Substituting
\[
p(f_1) = \frac{b_1}{f_1} + a_0 + a_1 f_1 + \cdots ~,
\] 
we get the following equation near $f_1\approx 0$ involving coefficients ${b_1},\,a_0$ and $a_1$:
\[
\frac{{b_1}^2-B^2}{f_1^2}+\frac{2 {a_0} {b_1}}{f_1}
-2 a B+{a_0}^2+2 {a_1} {b_1}-B \hbar +{b_1} \hbar +{E_n} = 0~.
\]
From above, we get $b_1=B,\, a_0=0$ and $a_1=\frac{2 a B-{E_n}}{2 B}$. Since $dx \rightarrow df_1/f_1^2$, only the $a_1$-term contributes in Eq. (\ref{eq.Q_Condition}).

Now let us analyze the singularity structure of $p(f_1)$ near $z=1/f_1 =0$.  The QHJ equation for this case is
\[
p^2+\hbar \, \frac{dp}{dz} + {E_n}
-\frac{a^2-a \hbar }{z^2}-2 a B-B^2 z^2-B \hbar =0~.
\]
Substituting $p(z)  = b_1/z + a_0 + a_1 z + \cdots$, near $z\approx 0$ we get 
\[
\frac{-a^2+a \hbar +{b_1}^2-{b_1} \hbar }{z^2}+ \frac{2 {a_0} {b_1}}{z}
-2 a B+{a_0}^2
+2 {a_1} {b_1}+{a_1} \hbar -B \hbar +{E_n} =0,
\]
which yields $b_1=a,\, a_0=0$ and $a_1=\frac{2 a B+B \hbar -{E_n}}{2 a+\hbar }$. 
In this case, since $dx = -dz/z^2$, again only the $a_1$-term contributes to Eq. (\ref{eq.Q_Condition}).

Putting the contributions from singularities at $f_1=0$ and $f_1=\infty$ together, we get 
\begin{equation}
	\frac{1}{2} \left[  2\pi i\,\hbar  \left(\frac{2 a B-{E_n}}{2 B}-a\right)\right] =2\pi i\,n \hbar~, \label{eq.ClassIIIA-condition}
\end{equation}
which, with the identification $B \to -\frac{\omega }{2}$,  leads to the spectrum for the radial oscillator:
\[
{E_n} = 2 n \omega  \hbar~,
\]
which agrees with Eq. (\ref{eq.Energy.ClassIII}).
The factor of $\frac12$ in Eq. (\ref{eq.ClassIIIA-condition}) is due to the fact that on the complex $f_1$-plane, we have moving poles on the negative and positive sides of the real axis, while only those on the positive real axis are physical, and hence they double count in the complex integration \cite{Leacock1,Leacock2,Kapoor1,Kapoor2}.
\subsubsection*{Class IIIB: $\lambda\ne 0$}
In this case we have  $W=af_1+f_2$, where $f_1$ and $f_2$ satisfy $f_1'=f_1^2-\lambda$, $f_2'=f_1\,f_2 -\varepsilon$. Solving
the second differential equation
\footnote{The homogeneous solution of $f_2'-f_1\,f_2 =\varepsilon$ is
	$f_2 =B \exp{\left[ \int f_1 \, dx\right]}  = B \exp{\left[ \int f_1 \, \frac{df_1}{f_1^2-\lambda}\right]}  =B \exp{\left[ \frac12  \int \frac{df_1^2}{f_1^2-\lambda}\right]}  = B \sqrt{f_1^2-\lambda} ~.
	$
	A particular solution is $\left(\frac{\varepsilon}{\lambda}\right)\, f_{1}$. Thus, the super potential can be written as $\left( a+\frac{\varepsilon}{\lambda}\right)  f_1 +   B \sqrt{f_1^2-\lambda} \equiv a f_1 +   B \sqrt{f_1^2-\lambda}~$.  Thus, we have effectively set $\varepsilon=0$.} 
for $f_2$, we get $f_2 = B \sqrt{f_1^2-\lambda}$.  Hence, the superpotential for  this case reduces to $W=a\, f_1+ B \sqrt{f_1^2-1}$. 

We first consider $\lambda>0$, which implies that we can set $\lambda=1$. Furthermore, from Eq. (\ref{eq.PDE1}) and the requirement that $dg/da>0$, we find $a<0$. 

To avoid working with the square root of $f_1$ in the superpotential, following \cite{Gangopadhyaya2020}, we define a function $y$ by
\[
y(x)=\sqrt{\frac{f_1(x)-1}{f_1(x)+1}}~,
\]
which gives $dy/dx=y$.  The function $f_1$ and the superpotential $W$ are then given by 
\[
f_1 = -\frac{y+y^{-1}}{y-y^{-1}} \qquad \mbox{and} \qquad W=-a~\left(\frac{y+y^{-1}}{y-y^{-1}}\right)+\frac{2 B}{y-y^{-1}}~.
\]
This change of variable that casts the superpotential as a function of both $y$ and $y^{-1}$ results in singularities at both $y=0$ and $y=\infty$, and thus the domain reduces to $y\in (0,\infty)$. Hence, on the complex  $y$-plane  all physical moving poles will necessarily be on the positive real $y$-axis. The QHJ condition of Eq. (\ref{eq.QHJ2}) becomes
\begin{eqnarray}
	p^2 &-& \hbar\,y\,\frac{dp}{dy} +{E_n} +
	\frac{
		-a^2+y \left(4 a B-2 B \hbar\right)+ y^2 \left(-2 a^2+4 a \hbar -4 B^2\right) 
	}
	{\left(y^2-1\right)^2}
	\nonumber\\
	&+&
	\frac{
		 y^3 (4 a B-2 B \hbar )-{a^2 y^4}
	}
	{\left(y^2-1\right)^2}=0 ~.\label{eq.QHJ.classIIIB3}
\end{eqnarray}
Thus, in addition to zero and infinity, we also have singularities at $y=\pm 1$.  We will need to compute residues at all four singular points.

 Near $y=0$, setting $p=\frac{{b_1}}{y}+{a_0}+{a_1} y+\cdots $, the Eq. (\ref{eq.QHJ.classIIIB3}) reduces to
\[
\frac{{b_1}^2}{y^2}
+\frac{2 {a_0} {b_1}+{b_1} \hbar }{y}
-a^2+{a_0}^2+2 {a_1} {b_1}+{E_n}=0, 
\]
and hence we have $b_1=0$ and $a_0= -\sqrt{a^2-{E_n}} $. 
\footnote{As $E_n\to0$, we have $p\to W$.  But $W \to a$ as $y\approx 0$, hence we must have $p\to a$.  Which implies that since $a<0$, we must choose $a_0= -\sqrt{a^2-{E_n}} $.} Since under $x\to y$, the differentials are connected by $dx\to dy/y$, only the $a_0$-term contributes.

 Near $y=1$, setting $u=y-1$ and $p = b_1/u + a_0 + a_1\,u+\cdots$, the Eq. (\ref{eq.QHJ.classIIIB3}) reduces to
\[
-\frac{(-a+B-{b_1}) (-a+B+{b_1}+\hbar )}{u^2}
+
\frac{\hbar  ({b_1}-2 (-a+B))-2 (-a+B)^2+2 {a_0} {b_1}}{u}
+{\cal{O}}\left( u^0\right) =0~,
\]
and hence we have $b_1=-a+B$. Since near $y=1$, the differentials are connected by $dx\to dy/y= dy$, only the $b_1$-term contributes.

Near $y=-1$, setting $u=y+1$ and $p = b_1/u + a_0 + a_1\,u+\cdots$, the Eq. (\ref{eq.QHJ.classIIIB3}) reduces to
\[
\frac{(a+B-{b_1}) (-a-B-{b_1}+\hbar)}{u^2}
+\frac{2 ((-a - B)^2 + a_0 b_1) + (-2 a - 2 B + b_1) \hbar}{u}
+{\cal{O}}\left( u^0\right) =0~,
\]
and hence we have $b_1=a+B$. Since near $y=-1$, the differentials are connected by $dx\to dy/y= - dy$, again only the $b_1$-term contributes.

Finally, we look for the contribution coming from the singularity at $|y|\to \infty$, for which we set $z=1/y$ and
$p =\frac{{b_1}}{z}+{a_0}+{a_1} z+\cdots $ in Eq. (\ref{eq.QHJ.classIIIB3}).  This gives
\[
\frac{{b_1}^2}{z^2}+\frac{{b_1} (2 {a_0}-\hbar )}{z}-a^2+{a_0}^2+2 {a_1} {b_1}+{E_n}=0~,
\]
which gives $b_1=0$ and $a_0=\sqrt{a^2-{E_n}}$. In this case, $dx \to dy/y \to -dz/z$, and hence only need the $a_0$-term.

Thus the net contribution of the four singular points is
\[
-\sqrt{a^2-{E_n}} + (-a+B)+ (-)(a+B) +(-)\sqrt{a^2-{E_n}} = -2 \left( a+\sqrt{a^2-{E_n}}\right) 
\] 
As stated earlier, in this problem all $n$ physical moving poles lie on the positive side of real axis on the complex $y$-plane. However, there are $n$ ``spurious" moving poles on the negative real axis as well, with each contributing a residue of $i\,\hbar$ to the right-hand-side of Eq. (\ref{eq.Q_Condition}). Hence, we have 
\[
2\pi \,i\,\left[ -2 \left( a+\sqrt{a^2-{E_n}}\right) \right]   = 2\pi\, i\, (2 n \hbar)~,
\]
which gives $ E_n = a^2-(a+n\hbar)^2
$. Identifying $a=-A$, we get the energy for hyperbolic version of both Scarf and P\"oschl-Teller potential:
\[
E_n = A^2-(A^2-n\hbar)^2~,
\]
which agrees with Eq. (\ref{eq.Energy.ClassIII}) for $\lambda=1$ and $\varepsilon=0$.

Following a similar procedure, we can show that the case of $\lambda=-1$ leads to the trigonometric P\"oschl-Teller potential with eigenvalues 
\[
E_n =(A^2+n\hbar)^2- A^2.
\]

\section*{Derivation of Eigenfunctions}
We have shown that the spectra for all conventional potentials can be derived from QHJ. 
Guided by Ref. \cite{Kapoor3,Yesiltas2008,Girard2015}, where authors derive eigenfunctions of several explicit potentials, we now show that we can also derive the eigenfunctions of  shape invariant conventional potentials directly from Eq. (\ref{eq.PDE2solution}). As an illustrative example, we derive eigenfunctions of a class IIB superpotential, which is described in sufficient detail so that it could be used to derive eigenfunctions for any of the other classes.  

Superpotentials of class IIB have the form $W=a f_1+B/a$, where $f_1$ satisfies 
\begin{equation}
	\frac{df_1}{dx} = f_1^2-\lambda~. \label{eq.f1}
\end{equation} 
Depending on values of $\lambda$ and signs of $a$, it produces different conventional potentials.  We will now consider the case \footnote{If $\lambda\ne 0$, with scaling of $f_1$ and $x$, we can make $|\lambda|=1$.  Thus, $\lambda$ takes only three values: 0, 1 and -1.} with $a<0$ and $\lambda=1$. The needed boundary conditions for $W$ at the left and the right boundaries also require that we have $B<0$.  

We have already seen that eigenenergies are given by 
\begin{equation}
	E_n = a^2- {(a+n \hbar )^2}+\frac{B^2}{a^2}-\frac{B^2}{(a+n \hbar )^2}~, \label{eq.ClassIIB2.spectrum}
\end{equation}
which we will soon use. Going forward, we also set $\hbar=1$. As stated earlier, the quantum momentum function satisfies 
\begin{equation} 
	p\,^2(x) - \,p^{\,\prime}(x) = W^2(x)-  W^\prime(x) - E \label{eq.Ref.QHJ1} ~.
\end{equation}
Since the superpotential $W$ is given in terms of $f_1$, and because $f_1$ is a monotonically decreasing function 
\footnote{In this case, we have $W^\prime = a f_1^\prime = a (f_1^2-1). $  The relation $f_1^\prime = (f_1^2-1)$ implies that $f_1$ can never be equal to $\pm1$ at any point as otherwise $f_1$ will be constant. Thus, $f_1^2-1$ must have a fixed sign, and hence  $W$ must be monotonic function of $x$. However, from unbroken supersymmetry, we must have  $W^\prime >0$ for this case.  So, choosing $a<0$ requires that $f_1^\prime$ be negative as well.
}, 
we do a change of variable from $x\to S = -f_1$ and the quantum momentum function $p(x) \to \tilde{p}(S)$ then satisfies
\begin{equation} 
	\tilde{p}\,^2(S) - \,\dot{\tilde{p}}(S)\, \frac{dS}{dx} =\left( -a S+ B/a\right)^2 -  \left( S^2-1\right)  - E \label{eq.Ref.QHJ2} ~,
\end{equation}
where the dot over ${\tilde{p}}$ denotes a differentiation with respect to $S$. Following Ref. \cite{Kapoor2}, we define a function $\chi$ by
\begin{equation}
	\chi = \frac{\tilde{p}}{F}-\frac{1}{2}~ \frac{\dot{F}}F~,
\end{equation}
where $F$ stands for $ \frac{dS}{dx}$.  In variables $\chi$ and $S$, the QHJ equation becomes
\begin{equation}
	\chi ^2-\hbar  \dot{\chi}+\frac{ 1}{4}\left(\frac{\dot F^2}{F^2}-\frac{2 \ddot F}{F}\right) -\left(  \frac{W^2-\hbar  W'-{E_n}}{F^2}\right) =0~.
\end{equation}
Substituting $W=a (-S) +B/a$, we arrive at
\begin{equation}
	\chi ^2-\hbar  \dot{\chi} +\left( -\frac{\frac{B^2}{a^2}+(a-1) a S^2+a-2 B S-{E_n}}{F^2}+\frac{S^2}{F^2}+\frac{1}{F}\right)
	= 0~. \label{eq.chiEquation}
\end{equation}
Our objective here is to solve the above equation for $\chi$ and from there to derive the eigenfunctions $\psi$.  Since the equation is now given in terms of $S$ and derivatives with respect to $S$, our solutions for $\chi$ and $\psi$ will be in terms of this variable. 

We first note that the function $\chi$ has $n$ moving-poles, each with a residue of $(-1)$. In addition to the moving poles, it also has fixed poles that are engendered by the structure of the superpotential $W$.  As we see from Eq. (\ref{eq.chiEquation}), the fixed singularities for $ \chi $ are at $S=\pm 1$.  Hence, we write the function $\chi$ as 
\begin{equation}
	\chi = -\sum_{k=1}^n \frac{1}{S-S_{k}}+ \frac{b_-}{S+1}+ \frac{b_+}{S-1}+C~,  \label{eq.chiEquation2}
\end{equation}
where we have explicitly stated all of moving \footnote{The term with a summation on the right-hand-side comes from the $n$ moving poles.} and fixed poles and $C$ is an analytic function on the complex $S$-plane.

As  $|S|\to \infty$, the superpotential $W>>\sqrt{E_n}$ and the quantum momentum function $p\to W$ and the  function $\chi$ goes to
\[
\chi~\to~\left( \frac{W}{F}-\frac2\,\frac{F'}{F}\right) = \left(\frac{\frac{B}{a}-a S}{1-S^2}+\frac{\hbar S}{1-S^2}\right)\to \frac{a-\hbar}{S}~.
\]
I.e., $\chi\to 0$ for large $|S|$. Hence, from the Liouville's theorem the analytic function $C$ is bounded and must be a constant.  From its value for large $|S|$, the constant $C$ should be zero. 

We substitute the expression for $\chi$ from Eq. (\ref{eq.chiEquation2}) into Eq. (\ref{eq.chiEquation}) and expand separately around $S=\pm 1$.  Near $S=1$ we get two possible values for the residue:
\[b_+=\frac{-1-\sqrt{a^2-2 B- {E_n}+\frac{B^2}{a^2}}}{2 } \quad \mbox{or}\quad \frac{-1+\sqrt{a^2-2 B- {E_n}+\frac{B^2}{a^2}}}{2 }~.
\]
Since, for the limit $E_n\to 0$, we must have $\chi\to \left( \frac{W}{F} + \cdots \right)=\frac{a-\frac{B}{a}}{2 u} +\cdots$, we must choose
\footnote{~In the limit $E_n\to 0$,  
	\[
	\frac{-1-\sqrt{a^2-2 B- {E_n}+\frac{B^2}{a^2}}}{2 a^2} 
	\to 
	\frac{1}{2} \left(-\sqrt{\frac{\left(a^2-B\right)^2}{a^2}}-1\right)=
	\frac{1}{2} \left(-\frac{B-a^2}{a}-1\right)=
	\frac{1}{2} \left(-\frac{B}{a}+a-1\right)~,
	\]
exactly as $\chi$ should behave near $ S = 1 $. }
$${b_+}=\frac{-1-\sqrt{a^2-2 B- {E_n}+\frac{B^2}{a^2}}}{2 }.$$  With a similar reasoning, we see that we must have ${b_-}=\frac{1}{2} \left(\frac{\sqrt{a^4+2 a^2 B-a^2 {E_n}+B^2}}{a}-1\right)$.  

At this point,  we substitute  $E_n = a^2- {(a+n )^2}+\frac{B^2}{a^2}-\frac{B^2}{(a+n)^2}$ that we derived using the singularity of QMF, see Eq. (\ref{eq.ClassIIB.alt0}), which gives 
$$b_\mp\,=\, \frac{1}{2} \left(\pm \, \frac{B}{a+n}+a+(n-1)\right).$$ 

The first term on the right-hand-side of Eq. (\ref{eq.chiEquation2}) can be written in terms of the logarithmic derivative of a $n$-th order polynomial $P=\Pi_{k=1}^n \left( {S-S_{k}}\right) $, where $S_{k}$ denotes the zeroes of the polynomial. Then, we can write $\sum_{k=1}^n \frac{1}{ {S-S_{k}}} = \frac{\dot{P}}{P}$, and hence we have 
\begin{equation}
	\chi = -\frac{\dot{P}}{P}+ \frac{b_-}{S+1}+ \frac{b_+}{S-1}~.  \label{eq.chiEquation3}
\end{equation}
Substituting this form of $\chi$ in Eq. (\ref{eq.chiEquation}), we get an equation for the polynomial $P\,$:
\begin{eqnarray}
\left(1-S^2\right) \ddot P+
 \left(2(a+n-1)\, S -\frac{2 B}{a+n}\right) \dot P
-n (2 a+n-1) P=0~.
\end{eqnarray}
Defining 
$$\alpha =\frac{B}{a+n}-a-n \qquad \mbox {and} \qquad\beta =-\frac{B}{a+n}-a-n, $$ 
the equation reduces to the familiar Jacobi differential equation:
\begin{equation}
	\left(1-S^2\right) \ddot{P}+
	\left(\beta-\alpha -(\alpha+\beta+2)S\right) \dot{P}+ 
	n (n+\alpha+\beta+1)\, P=0~,
\end{equation}
and its solutions are the Jacobi Polynomials $P_n^{(\alpha,\beta)}(S)$.  In terms of $\alpha$ and $\beta$, the function $\chi$ is now given by 
\begin{equation}
	\chi = -\frac{\dot{P_n}}{P_n}+ \frac{\frac{1}{2} (-\beta -1)}{S+1}+ \frac{\frac{1}{2} (-\alpha -1)}{S-1}~.  \label{eq.chiEquation4}
\end{equation}
\bigskip
The eigenfunctions are then given by 
\begin{eqnarray}
	\psi_n &\sim& e^{-\int p\,dx}\sim e^{-\int \tilde{p}\,dS/S^\prime} 
	=  e^{-\int \left( \chi +\frac{1}{2}\,\frac{\dot{F}}F \right) \,dS} \nonumber\\
	&\sim& 
	e^{\int \left( \frac{\dot{P_n}}{P_n}+ \frac{\frac{1}{2} (\beta +1)}{S+1}+ \frac{\frac{1}{2} (\alpha +1)}{S-1} -\frac{1}{2}~ \frac{2S}{S^2-1} \right) \,dS}\nonumber\\
	&\sim& 
	e^{\left( \log{P_n}+{\frac{1}{2} (\beta +1)}\log({S+1})+ {\frac{1}{2} (\alpha +1)}\log({S-1}) -\frac{1}{2}~ \log({S-1})-\frac{1}{2}~ \log({S+1}) \right) }\nonumber\\
	&=& N
	\left( 1+S\right)^{\frac{\beta}{2}} \, \left( 1-S\right)^{\frac{\alpha}{2} }\, P_n^{(\alpha,\beta)}(S)
\end{eqnarray}
For the specific case of $S=\tanh x, \, -a=A$ and $-B\to B$, the superpotential goes to $W = A \, \tanh x +B/A$, the superpotential for hyperbolic Rosen-Morse or Rosen-Morse II and the result matches with that of Ref. \cite{De}.

Thus, unlike in SUSYQM where ladder operators generate one eigenfunction at a time, in QHJ formalism eigenfunctions are determined in one-shot in a manner similar to what happens when we directly solve the schr\"odinger equation. Thus, while unbroken SUSY and shape invariance do make problems solvable in disparate formalisms, the end results do not always manifest the same way.

\begin{table}[h!]
\begin{center}	
	\begin{tabular}{||l|l|c|l||}
		\hline\hline
		\textbf{Class}& \textbf{Superpotential}&\textbf{Energy $E_n$ }&\qquad \textbf{Name}\\ 
		\hline
		IA &$\alpha\, a+f_2$ & $n \hbar \omega$ &1D-Harmonic \\
		$\alpha=0$& $\varepsilon = -\frac 12 \omega$&&Oscillator   \\
		\hline
		IB &$\alpha a+ f_2$&$a^2-\left(a+n\hbar \right) ^2$ &Morse\\
		$\alpha=-1$&$a<0\,;~\varepsilon =0$&& \\
		\hline
		IIA&$a \, f_1+B/a$ & $\frac{B^2}{a^2}-\frac{B^2}{(a+n \hbar )^2}$ & Coloumb\\ 
		$\lambda=0$& $a>0$&& \\
		\hline
		IIB1&$a \, f_1+B/a$		& ${(a+n \hbar )^2}-a^2$ 	& Rosen-Morse  \\
		$\lambda<0$& $a>0$ 	& $+ \frac{B^2}{a^2}-\frac{B^2}{(a+n \hbar )^2}~  $	& (Trigonometric)\\
		\hline
		IIB2&$a \, f_1+B/a$&  $a^2-{(a+n \hbar )^2}$ &Rosen-Morse \\
		$\lambda>0$ &$a<0,\, B<0$ &$+ \frac{B^2}{a^2}-\frac{B^2}{(a+n \hbar )^2}~  $& (Hyperbolic) \\
		\hline
		IIB3&$a \, f_1+B/a$& $a^2-{(a+n \hbar )^2}$    &Eckart\\
		$\lambda>0$&$a>0$&$+ \frac{B^2}{a^2}-\frac{B^2}{(a+n \hbar )^2}~  $& \\
		\hline
		IIIA&$a\,f_1+B/f_1$&$2\,n \hbar \omega$ & 3D-Oscillator\\
		$\lambda=0$&$a>0\,;~B=-\frac12 \omega$&&\\
		\hline
		IIIB1&$a\,f_1+B\sqrt{f_1^2+|\lambda|}$ & $\left(a+n\hbar \right) ^2-a^2$& Scarf  \\	
		($\lambda<0$)&$ a>0 $&& (Trigonometric)\\
		\hline
		IIIB2&$a\,f_1+B\sqrt{\lambda-f_1^2}$   &$a^2-\left(a+n\hbar \right) ^2$&Scarf \\
		($\lambda>0$, $f_1^2<\lambda$) &$a<0$&& (Hyperbolic)\\
		\hline
		IIIB3&$a\,f_1+B\sqrt{f_1^2-\lambda}$  &$a^2-\left(a+n\hbar \right) ^2$ &P\"oschl-Teller \\
		($\lambda>0$, $f_1^2>\lambda$) &$a<0$&& (Hyperbolic) \\
		\hline\hline
	\end{tabular}
\end{center}
	\caption{The three shape invariant classes, their energies and the corresponding conventional potentials}
	\label{table:conventionalSIP}
\end{table}

\newpage
\section*{Conclusion:} Within supersymmetric quantum mechanics, shape invariance and unbroken supersymmetry ensure the solvability of a quantum mechanical system and help determine its eigenvalues and eigenfunctions. In a recent work \cite{Rasinariu2013}, it was proved that the shape invariance can be utilized to determine spectrum of a quantum system in deformation quantization formalism as well. Since many authors have derived eigenspectra for an array of conventional potentials \cite{Leacock1,Leacock2,Kapoor1,Kapoor2,Kapoor3,KapoorBook,Ganguly2006,Yesiltas2008,Ocak2008,Yesiltas2010,Gharbi2013,Girard2015,Gu2016,Poveda-Cuevas2016,Schatz2018} using QHJ formalism on a case-by-case basis,  it begs the question whether shape invariance played any role in this process. 
In this paper, using the general form of conventional superpotentials as derived from the shape invariance condition \cite{Gangopadhyaya2008,Bougie2010,symmetry}, we show that shape invariance is indeed sufficient to guarantee the solvability of all conventional potentials in QHJ formalism. 

We also note several directions in which scope of this work could enhanced.  We have mainly investigated conventional superpotentials that do not have an explicit dependence on $\hbar$.  While some studies \cite{Ranjani2012,Ranjani2019} have been done  of the rational extensions \cite{Quesne2008,Quesne2009,Quesne2012a,Quesne2012b,Odake2009,Odake2010,Ramos2011,Sasaki2010} that explicitly depend on $\hbar$, these studies have been limited to the extensions that originate from some specific conventional potentials.  An explicitly shape invariance based analysis would be an improvement. Furthermore, our analysis also did not include other types of shape invariance, such as multiplicative and cyclic.  It  would be worthwhile to see if the work of Refs. \cite{Gangopadhyaya2008,Bougie2010,symmetry} could be extended for multiplicative and cyclic shape invariance to determine the general forms of the associated potentials, and if yes, whether they would be solvable with QHJ as well. 

\section*{Acknowledgments:}
We are very thankful to the anonymous referees for their insightful and immensely helpful comments.  We also thank Jonathan Bougie and Constantin Rasinariu as this work benefited from many discussions with them on the subject.


\begin{thebibliography}{99}
\bibitem{Infeld}
L.~Infeld and T.~E. Hull, The factorization method; Rev. Mod. Phys. \textbf{23} (1951)
21--68.
%
\bibitem{Miller}
W.~Miller~Jr, \textit{Lie Theory and Special Functions} (Mathematics in Science and
Engineering); Academic Press, New York, NY, USA, 1968.
%
\bibitem{Gendenshtein1}
L.~E. Gendenshtein, Derivation of exact spectra of the Schr\"odinger equation by
means of supersymmetry; JETP Lett. \textbf{38} (1983) 356--359.
%
\bibitem{Witten}
E.~Witten, Dynamical breaking of supersymmetry; Nucl. Phys. B185 (1981)
513--554.
%
\bibitem{Solomonson}
P.~Solomonson and J.~W. Van~Holten, Fermionic coordinates and supersymmetry in
quantum mechanics; Nucl. Phys. \textbf{B196} (1982) 509--531.
%
\bibitem{CooperFreedman}
F.~Cooper and B.~Freedman, Aspects of supersymmetric quantum mechanics; Ann. Phys.
\textbf{146} (1983) 262--288.
%
\bibitem{Styer2002} Daniel F. Styer, et al., Nine formulations of quantum mechanics, 
Am.  Jour. of Phys. 70 (2002) 288. doi: 10.1119/1.1445404
%
\bibitem{Rasinariu2013} C. Rasinariu, Shape invariance in phase space, Fortsch. Phys. 61 (2013) 4-19.
%
\bibitem{Leacock1}R.A. Leacock and M.J. Padgett, Hamilton-Jacobi Theory and the Quantum Action Variable;  Phys. Rev. Lett., {\bf  50} 3, 1983.
%
\bibitem{Leacock2}R.A. Leacock and M.J. Padgett, 
Hamilton-Jacobi/action-angle quantum mechanics; Phys. Rev. D {\bf  28} 2491, 1983; Phys. Rev. A
{\bf  33} 2775, 1986;
Am. J. Phys. {\bf 55}, 261, 1986.
\bibitem{Gozzi}E. Gozzi, classical and Quantum Adiabatic Invariants, Phys. Lett. B {\bf  165}, 351, 1985.
%
\bibitem{Kapoor1}R.S. Bhalla, A.K. Kapoor and P.K.Panigrahi,
Quantum Hamilton–Jacobi formalism and the bound state spectra;
Mod. Phys. Lett. {\bf  A 12} 295, 1997.
%
\bibitem{Kapoor2}R.S. Bhalla, A.K. Kapoor and P.K.Panigrahi, Exactness of the supersymmetric WKB approximation scheme;
Am. J. Phys. {\bf 65} 1187, 1997.
\bibitem{Ganguly2006} D. J. Fernandez C. and A. Ganguly,
Exactly solvable associated Lame potentials and supersymmetric transformations,
Annals. Phys. 322 (2007) 1143-1161. DOI:
https://doi.org/10.1016/j.aop.2006.07.011
%
\bibitem{Yesiltas2008} \"O. Ye\c{s}ilta\c{s} and B. Demircio\u{g}lu, Quantum Hamilton–Jacobi Approach to Two Dimensional Singular Oscillator,
Chinese Phys. Lett. 25 (2008) 1935 .
%
\bibitem{Ocak2008}\"O. Ye\c{s}ilta\c{s} and S.B. Ocak,  
The Generalized PT-Symmetric Sinh-Gordon Potential Solvable within Quantum Hamilton–Jacobi Formalism. Int J Theor Phys 47, 415–420 (2008). https://doi.org/10.1007/s10773-007-9462-7
%
\bibitem{Yesiltas2010} \"O. Ye\c{s}ilta\c{s}, The quantum effective mass Hamilton–Jacobi problem, Jour. Phys. A: Math. Theor. 43 (2010) 095305. 
DOI 10.1088/1751-8113/43/9/095305
%
\bibitem{Gharbi2013} A. Gharbi and A. Bouda,
Energy spectra of Hartmann and ring-shaped oscillator potentials using the quantum Hamilton–Jacobi formalism,
Phys. Scr. 88 (2013) 045007. DOI 10.1088/0031-8949/88/04/045007
%
\bibitem{Girard2015} M.F. Girard, Analytical Solutions of the Quantum Hamilton-Jacobi Equation
and Exact WKB-Like Representations of One-Dimensional Wave Functions, https://arxiv.org/abs/1512.01356
%
\bibitem{Gu2016}
X.Y. Gu, M. Zhang, J.Q. Sun, Exact Solutions of Non-Central Potentials, Mod. Phys. Lett. B 24 (2010) 1759-1767. DOI: 10.1142/S0217984910024134

\bibitem{Poveda-Cuevas2016}L.A. Poveda-Cuevas and F.J. Poveda-Cuevas, On the nodes of wave function and the quantum Hamilton-Jacobi solution,
https://archive.org/details/arxiv-1609.01198
%
\bibitem{Schatz2018}
K. Schatz, B. Friedrich, S. Becker, and B. Schmidt,
Symmetric tops in combined electric fields: Conditional quasisolvability via the quantum Hamilton-Jacobi theory,
Phys. Rev. A 97 (2018) 053417.
%
\bibitem{KapoorBook}
This book is an excellent source for understanding the QHJ formalism. It brings together information from many papers the authors have written on the subject. 

A. K. Kapoor, P.K. Panigrahi, S. Sree Ranjani, \textit{Quantum Hamilton-Jacobi Formalism} (SpringerBriefs in Physics) 1st ed. 2022.

%
\bibitem{Gangopadhyaya2007} C. Cherqui, Y. Binder, and  A. Gangopadhyaya, {Shape invariance and the exactness of the quantum Hamilton-Jacobi formalism}; Phys. Lett. {\bf A 372} (2008) 1406-1415.
%
\bibitem{Gangopadhyaya2008}Gangopadhyaya, A. and  Mallow, J.V.  Generating shape invariant potentials. {\it Int. J. Mod. Phys. A} {\bf 2008}, \emph{23}, 4959--4978.
%
\bibitem{Bougie2010}
J.~Bougie, A.~Gangopadhyaya and J.~V. Mallow, Generation of a complete set of additive shape-invariant potentials from an Euler equation; Phys. Rev. Lett.
(2010) 210402:1--210402:4.
%
\bibitem{symmetry}
J.~Bougie, A.~Gangopadhyaya, J.~V. Mallow and C.~Rasinariu, Supersymmetric quantum
mechanics and solvable models; Symmetry 4~(3) (2012) 452--473.	
%
\bibitem{Kapoor3}S.S. Ranjani, K.G. Geojo, A.K. Kapoor and P.K. Panigrahi, 
Bound State Wave Functions through the Quantum Hamilton-Jacobi Formalism, Mod. Phys. Lett. A. 19 (2004) 1457.
%
\bibitem{Dutt1993}R. Dutt, A. Gangopadhyaya, A. Khare, A. Pagnamenta and U. Sukhatme,
Solvable quantum mechanical examples with broken supersymmetry, Phys. Lett. A174 (1993) 363-367.
%
\bibitem{Gangopadhyaya2001}A. Gangopadhyaya, J.V. Mallow and U.P. Sukhatme, Broken supersymmetric shape invariant systems and their potential algebras, Phys. Lett. A 283 (2001) 279–284.
%
\bibitem{Gangopadhyaya2021}A. Gangopadhyaya, J. Bougie and C. Rasinariu, Exactness of Semiclassical Quantization Rule for Broken Supersymmetry,  Jour. Phys. A: Math. Theor. 54 (2021) 295206. DOI 10.1088/1751-8121/ac060a
%
\bibitem{Cooper-Khare-Sukhatme}
F.~Cooper, A.~Khare and U.~Sukhatme, \textit{Supersymmetry in Quantum Mechanics}, World Scientific, Singapore, 2001.
%
\bibitem{Gangopadhyaya-Mallow-Rasinariu}
A.~Gangopadhyaya, J.~Mallow and C.~Rasinariu, \textit{Supersymmetric Quantum Mechanics: An Introduction} (2nd ed.), World Scientific, Singapore, 2017.
%
\bibitem{Khare1993}A. Khare and U. Sukhatme, New Shape Invariant Potentials in Supersymmetric Quantum Mechanics, Jour. Phys. A 26 (1993) L901-L904. https://doi.org/10.1088/0305-4470/26/18/003
%
\bibitem{Barclay1993}D. Barclay, R.  Dutt, A. Gangopadhyaya, A. Khare, A. Pagnamenta, and U.P. Sukhatme, New Exactly Solvable Hamiltonians - Shape Invariance and Self-Similarity, Phys. Rev. A 48 (1993) 2786-2797.
http://dx.doi.org/10.1103/PhysRevA.48.2786
%
\bibitem{Gangopadhyaya1996}
A. Gangopadhyaya and U.P. Sukhatme, Potentials with Two Shifted Sets of Equally Spaced Eigenvalues and Their Calogero Spectrum, Phys. Lett. A 224 (1996) 5-14. doi:10.1016/S0375-9601(96)00807-9
%
\bibitem{Sukhatme1997} U.P. Sukhatme, C. Rasinariu and A. Khare, Cyclic shape invariant potentials, Phys. Lett. A 234 (1997) 401-409.  https://doi.org/10.1016/S0375-9601(97)00555-0
%
\bibitem{Gangopadhyaya1994} A. Gangopadhyaya, P.K. Panigrahi and U.P. Sukhathme,
Analysis of inverse-square potentials using supersymmetric quantum mechanics, J. Phys. A: Math. Gen. 27 (1994) 4295-4300.
%
\bibitem{Che2003}K.M. Cheng, P.T. Leung and C.S. Pang, Exactness of supersymmetric WKB method for
translational shape invariant potentials, Jour. Phys. A: Math. Gen. 36 (2003) 5045–5060.
%
\bibitem{Ramos99}J. F. Carinena and A. Ramos, Riccati Equation, Factorization Method and Shape Invariance, Rev. Math. Phys.; \textbf{12}, (2000) 1279--1304.
%
\bibitem{Ramos00}J. F. Carinena and A. Ramos, Shape-invariant potentials depending on $n$-parameters
transformed by translation; J. Phys. A: Math. Gen. \textbf{33} (2000) 3467--3481.
%
\bibitem{Dutt} R.~Dutt, A.~Khare and U.~Sukhatme, Supersymmetry, shape invariance and exactly
solvable potentials; Am. J. Phys. \textbf{56} (1988) 163--168.
%
\bibitem{Gangopadhyaya2020}A.~Gangopadhyaya, J.~V. Mallow, C.~Rasinariu and J. Bougie, Exactness of SWKB for shape invariant potentials, Phys. Lett. \textbf{A}384 (2020) 126722.
%
\bibitem{De} R. De, R. Dutt and U.P. Sukhatme, Mapping of shape invariant potentials under point canonical transformations, Jour. Phys. A 25 (1992) L843-850.
%
\bibitem{Ranjani2012}S. Sree Ranjani, P.K. Panigrahi P K, A.K. Kapoor, A. Khare and A. Gangopadhyaya 2012 Exceptional
orthogonal polynomials, QHJ formalism and SWKB quantization condition J. Phys. A: Math.
Theor. 45 055210 (arXiv:1009.1944)

\bibitem{Ranjani2019}S. Sree Ranjani, Quantum Hamilton–Jacobi route to exceptional Laguerre polynomials and the corresponding rational potentials, Pramana – J. Phys. (2019) 93:29, https://doi.org/10.1007/s12043-019-1787-2
%
\bibitem{Quesne2008}C. Quesne, Exceptional orthogonal polynomials, exactly solvable potentials and supersymmetry, J. Phys. A:
Math. Theor. 41 (2008), 392001, 6 pages, arXiv:0807.4087.
%
\bibitem{Quesne2009}C. Quesne, Solvable rational potentials and exceptional orthogonal polynomials in supersymmetric quantum
mechanics, SIGMA 5 (2009), 084, 24 pages, arXiv:0906.2331.
%
\bibitem{Quesne2012a}C. Quesne, Exceptional orthogonal polynomials and new exactly solvable potentials in quantum mechanics,
J. Phys. Conf. Ser. 380 (2012), 012016, 13 pages, arXiv:1111.6467.
%
\bibitem{Quesne2012b}C. Quesne, Novel Enlarged Shape Invariance Property and Exactly Solvable Rational Extensions of the Rosen–Morse II and Eckart Potentials, Symmetry, Integrability and Geometry: Methods and Applications SIGMA 8 (2012), 080.
%
\bibitem{Ramos2011}A. Ramos, On the new translational shape-invariant potentials, J. Phys. A: Math. Theor. 44 (2011), 342001, 9 pages, arXiv:1106.3732.
%
\bibitem{Odake2009}S. Odake, R. Sasaki, Infinitely many shape invariant potentials and new orthogonal polynomials, Phys. Lett. B 679 (2009), 414–417, arXiv:0906.0142.
%
\bibitem{Odake2010}S. Odake, R. Sasaki, Infinitely many shape-invariant potentials and cubic identities of the Laguerre and Jacobi polynomials, J. Math. Phys. 51 (2010), 053513, 9 pages, arXiv:0911.1585.
%
\bibitem{Sasaki2010}R. Sasaki, S. Tsujimoto, A. Zhedanov, Exceptional Laguerre and Jacobi polynomials and the corresponding
potentials through Darboux–Crum transformations, J. Phys. A: Math. Theor. 43 (2010), 315204, 20 pages,
arXiv:1004.4711.

\end{thebibliography}
\end{document}